\documentclass[]{spie}  

\usepackage{amsmath,amsfonts,amssymb}
\usepackage{graphicx}
\usepackage[colorlinks=true, allcolors=blue]{hyperref}
\usepackage{mathtools} 

\usepackage[english]{babel}
\usepackage{hyperref}

\title{Convolutional Neural Network Denoising in Fluorescence Lifetime Imaging Microscopy (FLIM)}

\author[1]{Varun Mannam$^*$}
\author[1,2]{Yide Zhang}
\author[1]{Xiaotong Yuan}
\author[3]{Takashi Hato}
\author[3]{Pierre C. Dagher} 
\author[4]{Evan L. Nichols}
\author[4]{Cody J. Smith}
\author[3]{Kenneth W. Dunn}
\author[1]{Scott Howard}

\affil[1]{Department of Electrical Engineering, University of Notre Dame, Notre Dame, IN 46556, USA}
\affil[2]{Caltech Optical Imaging Laboratory, Andrew and Peggy Cherng Department of Medical Engineering,  California Institute of Technology, Pasadena, CA 91125, USA}
\affil[3]{Department of Medicine, Division of Nephrology, Indiana University, Indianapolis, Indiana 46202, USA}
\affil[4]{Department of Biological Sciences, University of Notre Dame, Notre Dame, Indiana 46556, USA}
\authorinfo{*e-mail: vmannam@nd.edu}

\begin{document} 
\maketitle

\begin{abstract} 
Fluorescence lifetime imaging microscopy (FLIM) systems are limited by their slow processing speed, low signal-to-noise ratio (SNR), and expensive and challenging hardware setups. In this work, we demonstrate applying a denoising convolutional network to improve FLIM SNR. The network will integrated with an  instant FLIM system with fast data acquisition based on analog signal processing, high SNR using high-efficiency pulse-modulation, and cost-effective implementation utilizing off-the-shelf radio-frequency components. Our instant FLIM system simultaneously provides the intensity, lifetime, and phasor plots \textit{in vivo} and \textit{ex vivo}. By integrating image denoising using the trained deep learning model on the FLIM data, provide accurate FLIM phasor measurements are obtained. The enhanced phasor is then passed through the K-means clustering segmentation method, an unbiased and unsupervised machine learning technique to separate different fluorophores accurately. Our experimental \textit{in vivo} mouse kidney results indicate that introducing the deep learning image denoising model before the segmentation effectively removes the noise in the phasor compared to existing methods and provides clearer segments. Hence, the proposed deep learning-based workflow provides fast and accurate automatic segmentation of fluorescence images using instant FLIM. The denoising operation is effective for the segmentation if the FLIM measurements are noisy. The clustering can effectively enhance the detection of biological structures of interest in biomedical imaging applications.
\end{abstract}


\section{Introduction}
Fluorescence lifetime imaging microscopy (FLIM) is a powerful technique in biomedical research that provides enhanced molecular contrast in addition to conventional fluorescence imaging by measuring the fluorescence decay lifetime of excited fluorophores in optical microscopy \cite{mannam2020machine}. The fluorescence lifetime indicates ion or dissolved oxygen concentration, thus enabling FLIM to measure the micro-environment in biology \cite{zhang2016super, zhang2019three, zhang2019super}. However, traditional FLIM systems are limited by their slow processing speed, low signal-to-noise ratio (SNR), and expensive and challenging hardware setups \cite{mannam2020machine}. Previously, we have demonstrated a novel instant FLIM system \cite{zhang2020instant} based on analog signal processing for fast data acquisition as shown in Fig.~\ref{instant_flim}. This Instant FLIM system provides high SNR using high-efficiency pulse-modulation, and cost-effective implementation by utilizing off-the-shelf radio-frequency components like phase shifters, mixers, and low-pass filters (LPFs). Moreover, due to the analog processing in instant FLIM, intensity, lifetime, and phasor plots for 2D, 3D, or 4D imaging \textit{in vivo} and \textit{ex vivo} are measured simultaneously.

Machine learning (ML) has recently gained significant attention for its improved performance in image processing, especially in image denoising \cite{goodfellow2016deep}. Multiple ML methods have been successfully utilized to denoise images with Gaussian noise or Poisson noise or mixed Poisson-Gaussian noise \cite{zhang2019poisson}. However, ML methods normally require a separate training dataset for image denoising. This paper demonstrates phasor image denoising using ML models including Noise2Noise or DnCNN that are pre-trained on 12000 fluorescence intensity images. Using the pre-trained ML models, we eliminate the requirement for a new training dataset requirement and demonstrate accurate phasor denoising results with a much faster processing speed. Also, lifetime segmentation with unbiased ML technique by using the denoised phasor is explained in the next section. 

\section{Methods}
\begin{figure}[!t]
\centering
\includegraphics[page=1,width=1.0\linewidth]{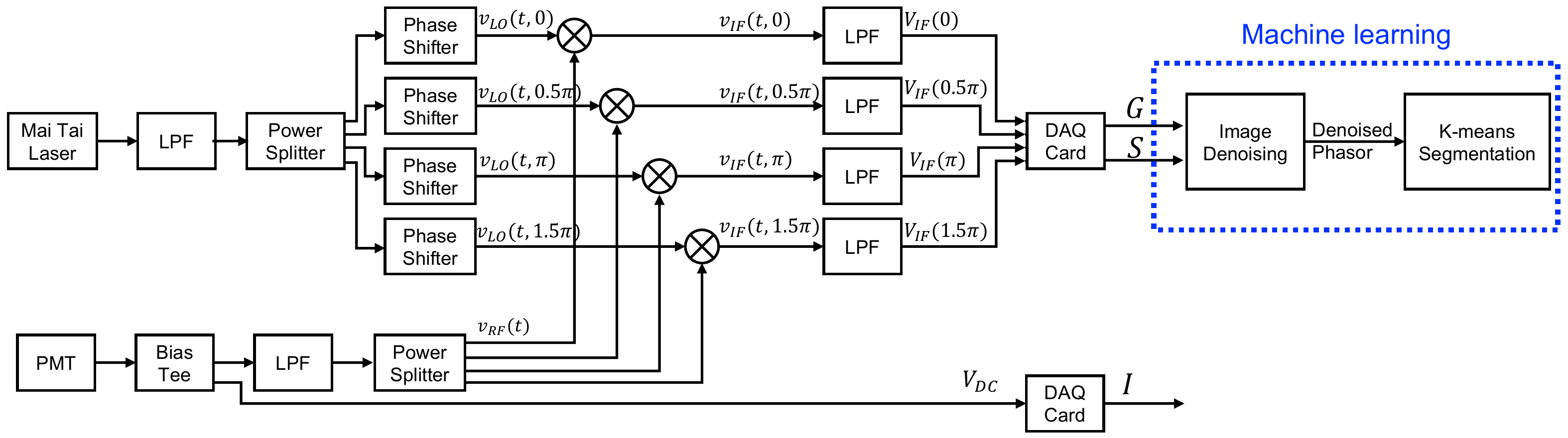}
\caption{Block diagram of the Instant FLIM system \cite{zhang2020instant}. Diagram shows how analog signals are processed in instant FLIM to extract the lifetime information.} \label{instant_flim}
\end{figure}

From Fig.~\ref{instant_flim}, the $G$ and $S$ images provided in the instant FLIM system are extracted by taking the difference between two mixers of complementary phases and are represented by the following equations.
\begin{equation}
    \begin{array}{l}
        S \propto V_\mathrm{IF}(0) - V_\mathrm{IF}(\pi), \\ 
        G \propto V_\mathrm{IF}(0.5\pi) - V_\mathrm{IF}(1.5\pi). 
    \end{array}
\end{equation}
The lifetime information is extracted by simply taking the ratio of $S/(\omega G)$ as shown below.
\begin{equation}
\tau = \frac{1}{\omega}\frac{V_{IF}(0) - V_{IF}(\pi)}{V_{IF}(0.5\pi) - V_{IF}(1.5\pi)}.
\end{equation}
where $\omega = 2\pi f_\mathrm{mod}$, and $f_\mathrm{mod}$ is the modulation frequency of the instant FLIM system ($f_\mathrm{mod}$ is 80MHz in our instant FLIM system). 

To simplify the interpretation and visualization of lifetime information, phasor plots can be used with TD- and FD-FLIM data. In the phasor approach, each pixel's fluorescence lifetime value is transformed into a point in the 2D phasor plot, where the coordinate $g$ (in $G$ image) is the $x$ coordinate, and $s$ (in $S$ image) is the $y$ coordinate. Since different fluorophores and excited state reactions can alter the phasor coordinates ($g$ and $s$) values, they can be resolved on the phasor plot. For time correlated single photon counting (TCSPC) data, the phasors can be acquired using the transformations $g_i = \int_0^\infty I(t)\cos(wt)dt/\int_0^\infty I(t)dt$ and $s_i = \int_0^\infty I(t)\sin(wt)dt/\int_0^\infty I(t)dt$, where $I(t)$ is the TCSPC data at the $i\mathrm{-th}$ pixel. For FD-FLIM data, the phasors can be calculated using $g_i = m_i\cos(\phi_i), s_i =  m_i\sin(\phi_i)$, where $m_i$ and $\phi_i$ are the modulation degree change and the phase shift of the emission with respect to the excitation at the $i\mathrm{-th}$ pixel. In the phasor plot, pixels with similar fluorescence decays converge to a region and form a cluster. This feature can be used to segment the image based on the fluorescence decays. 

\begin{figure}[!ht]
\centering
\includegraphics[page=6,width=0.8\linewidth]{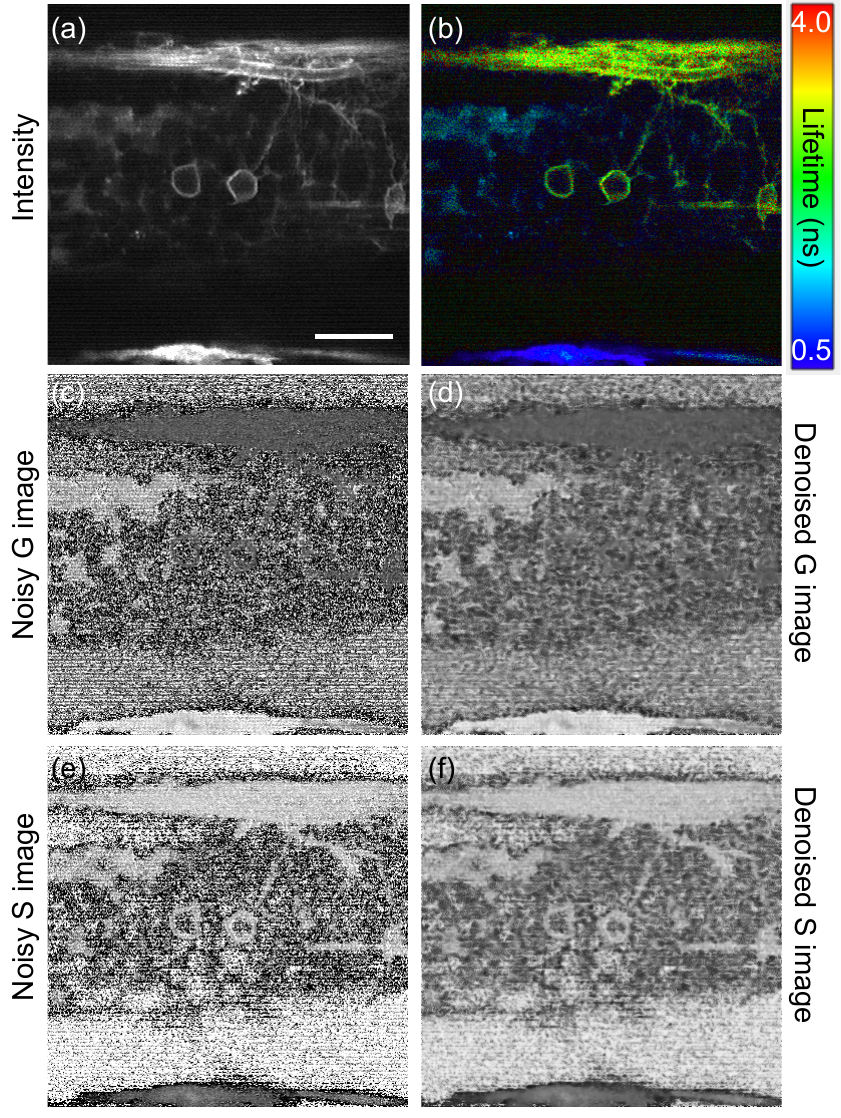}
\caption{The fluorescence intensity (a) and fluorescence composite lifetime (b) images of \textit{in vivo} zebrafish embryo (Tg (sox10:megfp) at 2 days post-fertilization) acquired with our custom-built instant FD-FLIM setup, respectively. Composite lifetime is the hue saturation value (HSV) representation of intensity and lifetime images together, where intensity and the fluorescence lifetimes are mapped to the pixels' brightness and hue, respectively. Noisy $G$ and $S$ images are shown in (c) and (e) and corresponding denoised $G$ and $S$ images are given in (d) and (f), respectively. The excitation laser was 800 nm with a power of 5.0 mW. 3D-volume of size $360 \times 360 \times 48$ where each slice depth of 1 $\mu$m with 48 slices and a pixel dwell time of 12 $\mu$s and was averaged three times for an improved signal-to-noise ratio. Scale bar: 20 $\mu$m.} \label{3D_ZF1_all}
\end{figure}

Fig.~\ref{3D_ZF1_all}(a) and (b) show the fluorescence intensity and composite lifetime of an \textit{in vivo} zebrafish embryo acquired with our instant FD-FLIM system in Fig.~\ref{instant_flim}. Here we show a single plane of the 3D-volume stack, where the neurons are marked with enhanced green fluorescent protein (EGFP), and the low-lifetime section ($\approx$0.5 ns) indicates the lateral nerve in the spinal cord. The phasor plots are acquired as a 2D grid image where $G$ and $S$ values are mapped to the x-axis and y-axis, respectively. More details about the phasor plots are provided in our review paper \cite{mannam2020machine}. Typically, both FLIM measurements and the generated phasor are noisy, which could lead to incorrect clustering of the biological sample boundaries. Conventional filtering approaches such as mean and median filters can process both $G$ and $S$ images to reduce the noise in phasors. However, they need to be used multiple times to reduce the significant amount of noise to get a good SNR. While \textbf{median filtering} preserves the edges in the denoised image compared to mean filtering \cite{digman2014phasor}, it needs to be used multiple times on $G$ and $S$ images to get clean phasors. Here, we demonstrate this behavior by showing the zebrafish 3D volume stack's phasor and applying the median filter multiple times on the $G$ and $S$ images. 

\begin{figure}[!ht]
\centering
\includegraphics[page=5,width=0.8\linewidth]{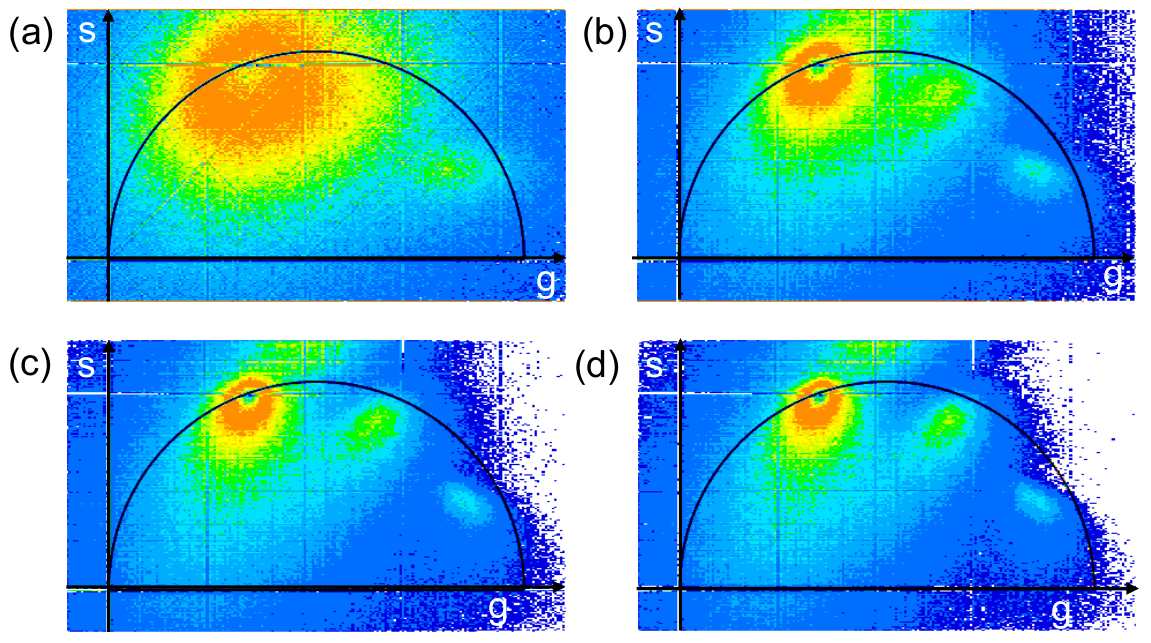}
\caption{The (a) raw phasor, (b) median filter applied once, (c) median filter applied twice, and (d) median filter applied three-times on the $G$ and $S$ images, respectively of the above mentioned \textit{in vivo} zebrafish embryo (Tg (sox10:megfp) at 2 days post-fertilization) acquired with our custom-built instant FD-FLIM setup.} \label{3D_ZF1_all_phasors}
\end{figure}

Fig.~\ref{3D_ZF1_all_phasors}(a) shows the noisy phasor of the zebrafish 3D-stack. Fig.~\ref{3D_ZF1_all_phasors}(b), (c) and (d) show the median filter \cite{digman2014phasor} being applied once, twice, and three times to reduce phasor noise respectively on the $G$ and $S$ images. The phasor in Fig.~\ref{3D_ZF1_all_phasors}(a) is unable to show the distribution of three different lifetime clusters corresponding to the lateral nerve ($\approx$ 0.5 ns), spinal cord ($\approx$ 1.5 ns), and EGFP neurons ($\approx$ 2.5 ns) due to noisy measurements. In contrast, after applying a median filter on the $G$ and $S$ images, the separation of the three clusters can be seen in the phasor plot. Applying median filtering multiple times (twice or three times) on the $G$ and $S$ images results in three fluorophore clusters in the phasor. Applying median filter more than three times, however, does not improve the phasor further and is computationally expensive. Table.~\ref{median_execution_time} shows the execution time of the median filtering on the 3D stacks.  

\begin{table}[]
\center
\begin{tabular}{|c|c|c|c|}
\hline
\multicolumn{1}{|l|}{} & \multicolumn{3}{c|}{Execution time (seconds)}                             \\ \hline
Stack name             & Median filter (once) & Median filter (twice) & Median filter (three-times) \\ \hline
G\_stack               & 0.4111                 & 0.6983                  & 0.9364                 \\ \hline
S\_stack               & 0.4107                 & 0.6931                  & 0.9160                 \\ \hline
\end{tabular}
\caption{Execution time for the median filtering on the $G$ and $S$ 3D-volume stack using the MATLAB tool \cite{MATLAB:2019}. \label{median_execution_time}}
\end{table}

Applying the median filter (multiple times) on the phasor obtained from noisy FLIM measurements is computationally expensive and inaccurate in identifying the proper fluorophore boundaries and corresponding lifetime values. Here, we demonstrate a novel pre-trained deep-learning model (using fluorescence intensity images) based on convolutional neural networks (CNNs) to instantly denoise the $G$ and $S$ images in ImageJ. The CNN network is similar to a typical auto encoder (AE) structure, which contains an encoder followed by a decoder block \cite{mannam2020performance}. In the encoder block, the noisy input image is reduced in shape by selecting only the image's essential features (rejecting noise). A noise-free image (from the encoder) is restored to the original noisy image shape in the decoder block. After training this CNN model with a large fluorescence microscopy images dataset, the $G$ and $S$ noisy images are denoised in the inference stage (test stage). The plugin is presented in our previous paper where more details about the CNN training and testing results are available \cite{mannam2020instant}.

\begin{figure}[!ht]
\centering
\includegraphics[page=4,width=0.8\linewidth]{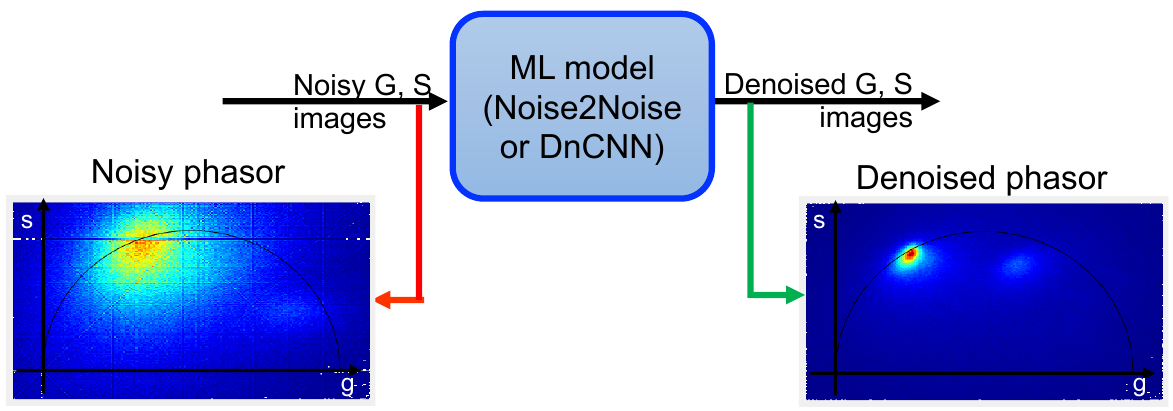}
\caption{Flow chart of our proposed method with 3D volume stack as shown in the Figure ~\ref{3D_ZF1_all} phasor before and after denoising (noisy and denoised phasors) using our ImageJ plugin \cite{mannam2020instant}.} \label{phasors_dncnn}
\end{figure}

Fig.~\ref{phasors_dncnn} shows the flow chart of our proposed method where applying the CNN based image denoising on the $G$ and $S$ images and extracting the phasor using the denoised images. Our proposed method also includes pre-processing (limits the range of $G$ and $S$ images) and post-processing (to convert back to original scale). Fig.~\ref{3D_ZF1_all}(c) and (d) show the noisy and denoised $G$ images of one of the slices in the 3D-stack, respectively. Similarly, Fig.~\ref{3D_ZF1_all}(e) and (f) show the noisy and denoised $S$ images of one of the slice in the 3D stack, respectively. Fig.~\ref{phasors_dncnn} shows the enhanced phasor when the noisy FLIM measurements are passed through deep learning-based image denoising.

Segmenting the denoised phasor representing multiple fluorophores with accurate lifetime values is the next critical step to show individual fluorophores. In the intravital imaging where autofluorescence is predominant, the user does not have prior knowledge of the fluorophores' locations and their corresponding lifetimes. Phasors with similar fluorescence decays appear to congregate and form a simple cluster in a phasor plot. Segmenting the phasor by selecting this cluster represents the similar fluorophore decays (or fluorophore lifetimes) and manually label these clusters with certain colors. The images extracted with each color show the respective fluorophores. Typically, the segmentation is performed by selecting the cluster as a region of interest (ROI) in the phasor, and lifetime information shows the corresponding fluorophore forms this cluster. However, this approach requires multiple trials and leads to different results every time depending on the ROIs selected by the user. Hence, to reduce the biased segmentation results, we presented a novel and unbiased approach for the automatic phasor labeling process using an unsupervised machine learning technique, i.e., K-means clustering \cite{zhang2019automatic}. This algorithm was developed by finding the $K$ centroids on the phasors plot and selecting the corresponding region around each centroid with a radius value \cite{macqueen1967some, jain2010data}. Our K-means clustering algorithm on the denoised phasor can automatically organize the phasor into accurate clustering. This segmentation result is more reliable than the clustering using the selected ROI's as it is unbiased. Segmentation results using this denoised phasor are explained in the next section.

\begin{figure}[!t]
\centering
\includegraphics[page=7,width=0.77\linewidth]{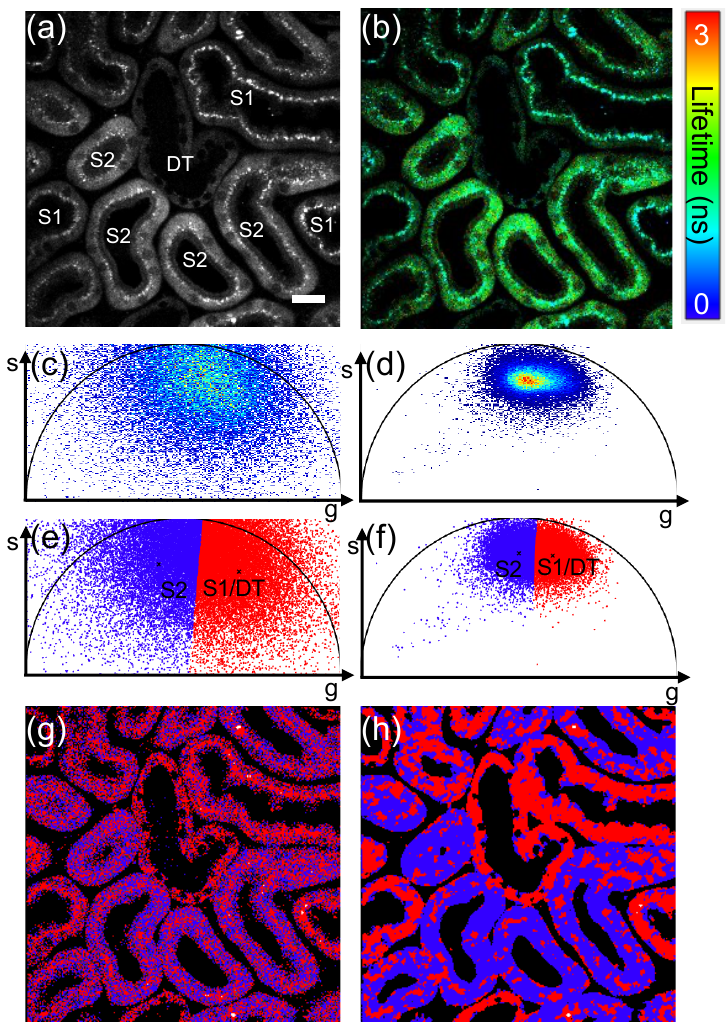}
\caption{K-means clustering segmentation applied to a living mouse kidney's phasor data acquired with a commercial FD-FLIM system \cite{zhang2019automatic}. Two-photon intensity image along with microtubules marking either $S1/DT$ or $S2$ (a), fluorescence lifetime image (b), noisy and denoised using our demonstrated method phasor plot (c) and (d), K-means clustering on phasors (e) and (f) before and after image denoising respectively. The $S1/DT$ and $S2$ microtubules have a short lifetime and long lifetime (as shown in red and blue colors). The overlap of both segments (segment1) $S1/DT$ and (segment2) $S2$ using noisy phasor and denoised phasors in (g) and (h), respectively. Scale bar: 20 $\mu$m.} \label{phasor_mouse_kidney}
\end{figure}

\section{Results}
We extended our CNN-based image denoising method to show the results with another commercial FLIM setup. Image denoising and segmentation are demonstrated in Fig.~\ref{phasor_mouse_kidney}(a) and (b) which show the intensity and fluorescence lifetime, respectively, for an \textit{in vivo} mouse kidney (male C57BL/6J mice at 8–10 weeks of age, obtained from The Jackson Laboratory) acquired with a commercial digital FD-FLIM system \cite{zhang2019automatic}. We labeled the intensity image with the mouse distal tubules ($DT$) and proximal tubules (upstream $S1$ and downstream $S2$) with distinct metabolic signatures, which can be resolved with FLIM phasors. Image denoising on the $G$ and $S$ images are performed using our demonstrated CNN based image denoising to extract the accurate phasor. Fig.~\ref{phasor_mouse_kidney}(c) shows the phasor labeled of the two clusters into a single image generated from the noisy phasor, and (d) shows the denoised phasor, which has high SNR by reducing the noise in the phasor (c). Furthermore, $S1$ and $DT$ are remarkably similar in FLIM lifetime signatures despite their morphological difference; therefore, we consider only categorizing $S1/DT$ and $S2$ as two clusters. Fig.~\ref{phasor_mouse_kidney}(e) shows an unbiased and unsupervised K-means clustering segmentation method. The K-means clustering results from the noisy phasor, where the red and blue color pixels in the phasor indicate the $S1/DT$ and $S2$ microtubules. The upstream proximal tubules have a shorter lifetime (right side of the noisy phasor) than the downstream tubules, which have a longer lifetime (left side of the noisy phasor). From the overlapping clusters in Fig.~\ref{phasor_mouse_kidney}(g), it is hard to accurately identify these tubules.

To address this issue, we use deep-learning based image denoising on the FLIM measurements ($G$ and $S$ images) to enhance the phasor, as shown in Fig.~\ref{phasor_mouse_kidney}(d). K-means clustering on the denoised phasor, shown in Fig.~\ref{phasor_mouse_kidney}(f), provides accurate segmentation results. From the segmented images after denoising, the phasor labeled image is shown in Fig.~\ref{phasor_mouse_kidney}(h), where the clear indication of dominant red and blue color tubules are mapped to $S1/DT$ and $S2$ proximal tubules, respectively. Hence, the demonstrated deep learning-based workflow provides fast and accurate automatic segmentation of fluorescence images acquired with the FLIM setup.

In summary, for the FLIM measurements, image denoising (using CNN based image denoising) and segmentation (K-means clustering) methods are beneficial if the FLIM measurements are noisy. The clustering can effectively enhance the detection of biological structures of interest in biomedical imaging applications (for example, accurate detection of the microtubules as shown in the Fig.~\ref{phasor_mouse_kidney}). The results used in this manuscript are provided in open-source and can be accessed via GitHub\footnote{\url{https://github.com/ND-HowardGroup/SPIE-CNN-FLIM-Denoising.git}}.

\section{Conclusion}
Fluorescence lifetime imaging microscopy (FLIM) measures the fluorophore decay rate by providing additional molecular contrast but limited by slow, low signal-to-noise ratio and costly setup. To overcome these limitations, we have demonstrated a novel high-speed, high-SNR FLIM system implementing real-time signal processing with off-the-shelf components. Instant FLIM provides the intensity, lifetime, and phasors for 2D, 3D, or 4D \textit{in vivo} imaging. We integrate image denoising using ML based trained model, which removes in the phasor noise followed by the K-means clustering segmentation for clear segments. The image denoising is useful for the noisy FLIM measurements that segment biological structures of interest.

\section*{Disclosures}
\noindent The authors declare no conflicts of interest.

\section*{Funding.}
This material is based upon work supported by the National Science Foundation (NSF) under Grant No. CBET-1554516. 

\acknowledgments 
Yide Zhang’s research was supported by the Berry Family Foundation Graduate Fellowship of Advanced Diagnostics $\&$ Therapeutics (AD$\&$T), University of Notre Dame. The authors further acknowledge the Notre Dame Center for Research Computing (CRC) for providing the Nvidia GeForce GTX 1080-Ti GPU resources for training the Fluorescence Microscopy Denoising (FMD) dataset\footnote{\url{https://curate.nd.edu/show/f4752f78z6t}} in TensorFlow.

\bibliography{report} 
\bibliographystyle{spiebib} 

\end{document}